\documentclass{article}

\usepackage{authblk}

\usepackage{siunitx}
\usepackage{amsmath}

\usepackage[sort,square,numbers]{natbib}

\usepackage{url}

\usepackage{graphicx}

\begin{document}

\title{\textbf{High-precision local transfer of van der Waals materials on nanophotonic structures}}
\author[1]{David Rosser}
\author[2]{Taylor Fryett}
\author[2]{Abhi Saxena}
\author[2]{Albert Ryou}
\author[1,2,*]{Arka Majumdar}
\affil[1]{Department of Physics, University of Washington 98195, USA}
\affil[2]{Department of Electrical and Computer Engineering, University of Washington, Seattle, Washinton 98195, USA}
\affil[*]{Corresponding Author: arka@uw.edu}

\maketitle

\begin{abstract}
Prototyping of van der Waals materials on dense nanophotonic devices requires high-precision monolayer discrimination to avoid bulk material contamination. We use the glass transition temperature of polycarbonate, used in the standard dry transfer process, to draw an in situ point for the precise pickup of two-dimensional materials. We transfer transition metal dichalcogenide monolayers onto a large-area silicon nitride spiral waveguide and silicon nitride ring resonators to demonstrate the high-precision contamination-free nature of the modified dry transfer method. Our improved local transfer technique is a necessary step for the deterministic integration of high-quality van der Waals materials onto nanocavities for the exploration of few-photon nonlinear optics on a high-throughput, nanofabrication-compatible platform.
\end{abstract}

\section{Introduction}
Atomically thin van der Waals (vdW) materials have generated strong interest in recent years for their possible electronic and optoelectronic applications \cite{liu_van_2016, liu_van_2019, xia_two-dimensional_2014}. The appeal of vdW materials for use as an active or passive material in low-loss nanophotonic devices hinges on their layered nature, which allows them to be integrated without concern for lattice-matching to the underlying substrate material \cite{fryett_cavity_2018}. The integration of vdW materials can thus be made independent of the device fabrication. The devices can be manufactured separately using existing high-throughput nanofabrication, including CMOS processes, and then the vdW material can be transferred on this pre-fabricated photonic platform to add new functionalities. The variety of vdW materials available with different optoelectronic properties provides for broad opportunities in the fabrication of light sources \cite{wu_monolayer_2015, ye_monolayer_2015}, modulators \cite{phare_graphene_2015}, detectors \cite{youngblood_waveguide-integrated_2015}, and nonlinear optical devices \cite{fryett_cavity_2018}.

Mechanically exfoliated and small-area chemical vapor deposition (CVD) grown vdW materials are pervasive in laboratory experiments due to their high quality and ease of device integration \cite{tan_recent_2017, cai_chemical_2018}. Various transfer techniques have been devised to facilitate rapid prototyping of vdW material heterostructures assembled from randomly located, micron-sized samples that are often surrounded by unwanted bulk material \cite{wang_one-dimensional_2013, lotsch_vertical_2015, hemnani_2d_2018, kim_van_2016}. For pure material studies the surrounding bulk materials do not pose a serious problem because there are no extended structures to avoid in the transfer process. In the realm of nanophotonics, however, stray bulk material can modify the optical properties of the structure under study. Moreover, many of these contaminants cannot be removed easily via etching or cleaning in solvents, often leading to ruined chips. Hence, a local transfer technique with improved monolayer discrimination is desired for high-yield vdW material integrated nanophotonic structures.

In this paper we demonstrate a modified polycarbonate-polydimethylsiloxane (PC-PDMS) transfer technique, which allows precise pickup and placement of vdW materials onto nanophotonic structures. As mentioned in Kim et al \cite{kim_van_2016}, the contact area (i.e. the region of the PC film which is in contact with the substrate) of the standard dome stamp method is limited to an approximately 50 \si{\micro \meter} x 50 \si{\micro \meter} area. The contact area of our process can be two orders of magnitude smaller than the dome method. We demonstrate the efficacy of our transfer process by placing $\text{WSe}_2$ onto a large-area silicon nitride spiral \cite{chen_design_2014} and two different semiconductor monolayers ($\text{WSe}_2$, $\text{MoSe}_2$) onto neighboring silicon nitride ring resonators \cite{bogaerts_silicon_2012}. Using this method, we have successfully transferred monolayer flakes as small as 4 \si{\micro \meter^2} and within 1 \si{\micro \meter} of bulk material, which is ostensibly limited only by the diffraction of light. 

\section{Experimental procedure}

\begin{figure}
	\centering
	\includegraphics[width=1.0\linewidth]{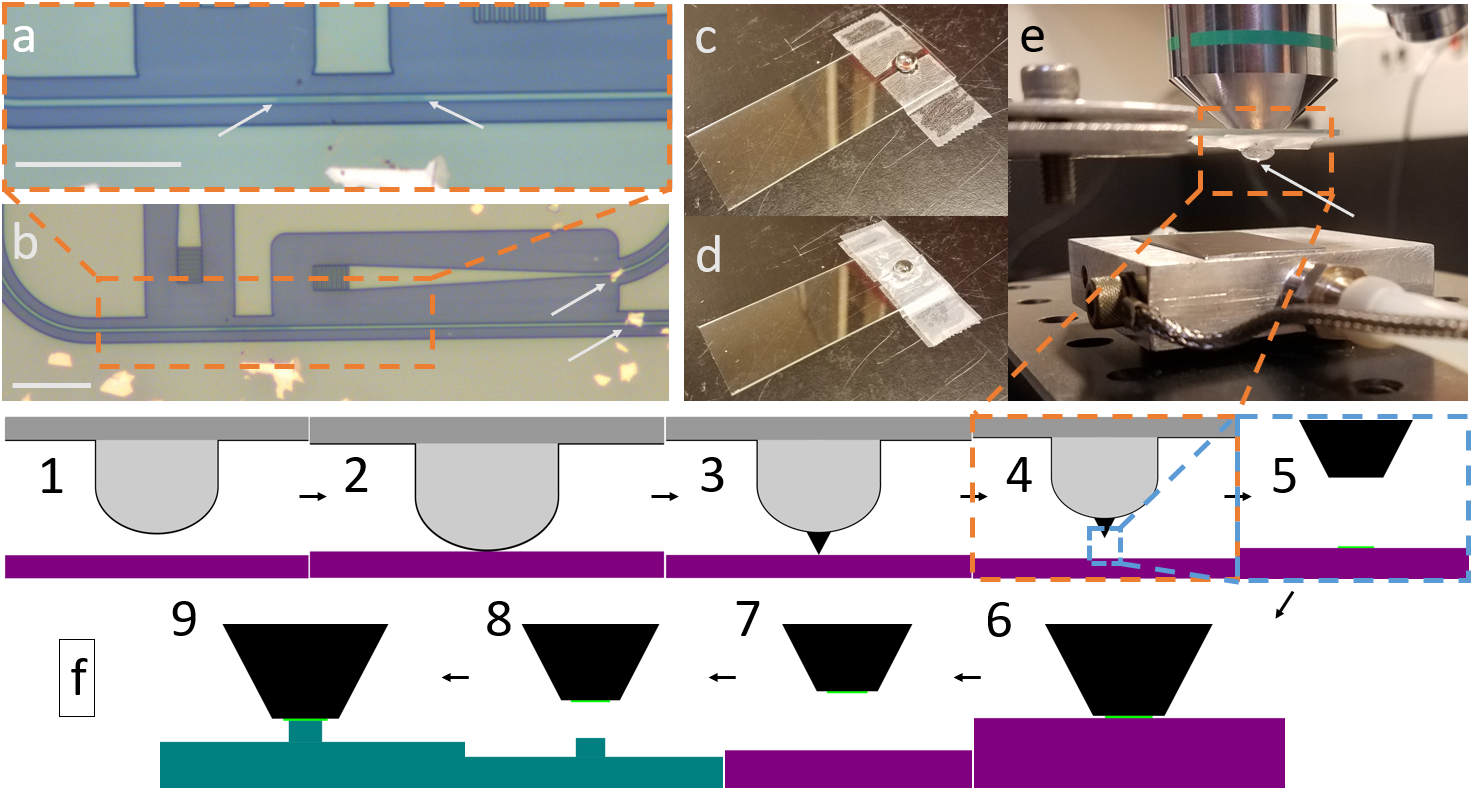}
	\caption{a) 2D material transferred onto a nanobeam cavity indicated by the arrows. The monolayer material is not visible on the SiN subtrate. b) Bulk material on waveguides indicated by the arrows. Scale bars are 10 \si{\micro\meter}. c) Dome stamp on a glass slide. d) PC film secured to the dome stamp with Scotch® tape. e) $\approx 1$ \si{\milli\meter} drawn PC point indicated by the arrow. f) Visual schematic of the procedure described in the text. Steps numbered 1-9. Purple is the stage substrate (e.g. SU-8 or $\text{SiO}_2$) and teal is the SiN waveguide. Dark gray is the glass slide, light gray is the PDMS dome, and black is the PC film. Green is the vdW material.}
	\label{fig:schematic}
\end{figure}

We first demonstrate the dome transfer method with a zoomed-in nanobeam cavity (Fig. \ref{fig:schematic}(a) and Fig. \ref{fig:schematic}(b)) to illustrate the scale of a nanophotonic device \cite{fryett_encapsulated_2018}. The monolayer is transferred on the nanobeam cavity resonator, but the monolayer is invisible under an optical microscope due to the poor optical contrast on the silicon nitride substrate. The dome stamp contaminates the nanophotonic devices with bulk material and tape residue which can significantly alter the transmission properties of the devices, sometimes to an extent where no transmission can be measured. Our local transfer method, which we describe below, allows for the precise pickup and placement of vdW material samples without the usual accompanying bulk material pieces. The local transfer method is demonstrated in a supplementary video.

\begin{figure}
	\centering
	\includegraphics[width=1.0\linewidth]{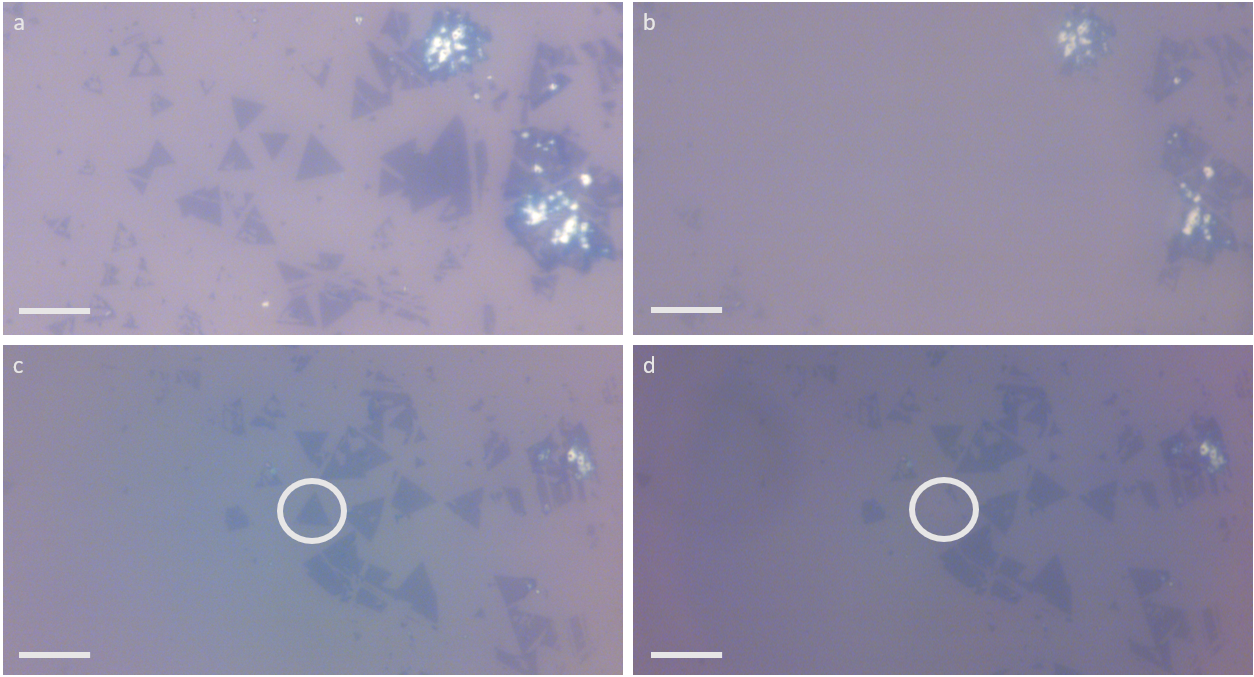}
	\caption{a) Before pickup - CVD grown 2D material on $\text{SiO}_2$. b) After pickup with PC dome - CVD grown 2D material on $\text{SiO}_2$. Any materials picked up will be deposited onto the nanophotonic device. c) Before pickup – CVD-grown 2D material on $\text{SiO}_2$. d) After pickup with PC point – CVD-grown 2D material on $\text{SiO}_2$ without the removed monolayer $\text{WSe}_2$ triangle. A small sliver of the monolayer's edge is left behind. Scale bars are 10 \si{\micro\meter}.}
	\label{fig:contactarea}
\end{figure}

The hemispherical dome stamp fabrication begins by preparing a 2-3 \si{\milli \meter} layer of cured PDMS (SYLGARD™184 Silicone Elastomer) cut into 6 \si{\milli \meter} diameter rounds. A second batch of PDMS is mixed from the silicone elastomer base with the curing agent, and placed in vacuum for 20 minutes for degassing. The liquid PDMS is then pipetted onto the round layer to form a hemisphere under the surface tension of the liquid. The domes are cured by leaving them in vacuum for 24 hours (Fig. \ref{fig:schematic}(c)).

The PC film (Sigma Aldrich® Poly(Bisphenol A carbonate), 7\% solution in chloroform) is secured to the hemispherical PDMS stamp using Scotch® tape with a hole punched into it as a window (Fig. \ref{fig:schematic}(d)). The sample stage is first set to 125 \si{\celsius} (Fig. \ref{fig:schematic}(f)-1) and always under vacuum to avoid picking up the chip. Under an optical microscope, the dome stamp is lowered into minimal contact with the sample stage (Fig \ref{fig:schematic}(f)-2). We use a SU-8 chip with pillars of varying diameters for the sample stage as a visual reference in the point formation. The dome is offset from the pillar, so it does not interfere with the melting PC. The SU-8 chip is not essential for the PC point formation. It is solely a pragmatic solution to making a point with the same diameter as the monolayer sample to prevent picking up additional material. The sample stage is then heated to 160 \si{\celsius}. After the stage equilibrates to the new temperature, the sample stage temperature is again set to 125 \si{\celsius}. As the sample stage decreases towards the lower temperature, the dome stamp is drawn away from the sample stage to separate the PDMS stamp from the PC film, which will still be adhered to the sample stage. The dome stamp is continuously pulled away from the substrate as a point is drawn in the PC film commensurate with the monolayer sample (Fig. \ref{fig:schematic}(f)-3). The point should be formed before the sample stage reaches the polycarbonate glass transition temperature (147 \si{\celsius}). It is imperative to intentionally pull the newly formed point away from the stage after the sample stage crosses the glass transition temperature (Fig. \ref{fig:schematic}(e) and Fig. \ref{fig:schematic}(f)-4).

During pickup of the monolayer we need to ensure that the monolayer sample is centered on the microscope objective along with the newly formed point (Fig. \ref{fig:schematic}(f)-5). As the hemispherical PDMS dome itself acts as a lens, the heated stage position has to be adjusted to maintain the monolayer sample in the focal plane of the objective. The point will manifest as a white disk. Pickup is performed by contacting the point to the monolayer (Fig. \ref{fig:schematic}(f)-6 and Fig. \ref{fig:schematic}(f)-7).

Finally, to transfer the monolayer onto a nanophotonic device the point is again brought close to the surface (Fig. \ref{fig:schematic}(f)-8). Due to the suspended nature of the PC point, melting can cause the point to droop unpredictably. For precise placement of the monolayer it is easiest to rapidly lower the PDMS dome stamp into contact with the monolayer to anchor it to the sample substrate (Fig. \ref{fig:schematic}(f)-9). The temperature of the sample stage is then raised to 180 \si{\celsius} to detach the PC as a sacrificial layer from the PDMS stamp. The PC film is dissolved in chloroform for 12 hours followed by a 30 minute isopropanol bath.

The main limitation of the dome stamp is that as we lower the PC dome to a close enough distance that we start to see Newton’s rings, the dome will suddenly contact the substrate with the previously mentioned 50 \si{\micro\meter} x \si{\micro\meter} area contact area (Fig. \ref{fig:contactarea}(a) and Fig. \ref{fig:contactarea}(b)). Anything in contact with the PC film will likely be picked up and transferred onto the nanophotonic device. By using the described local transfer method, we can pick up, for example, a single 10 \si{\micro\meter}$^2$ triangle of CVD-grown monolayer WSe2 heavily surrounded by unwanted material (Fig. \ref{fig:contactarea}(c) and Fig. \ref{fig:contactarea}(d)).

\begin{figure}
	\centering
	\includegraphics[width=1.0\linewidth]{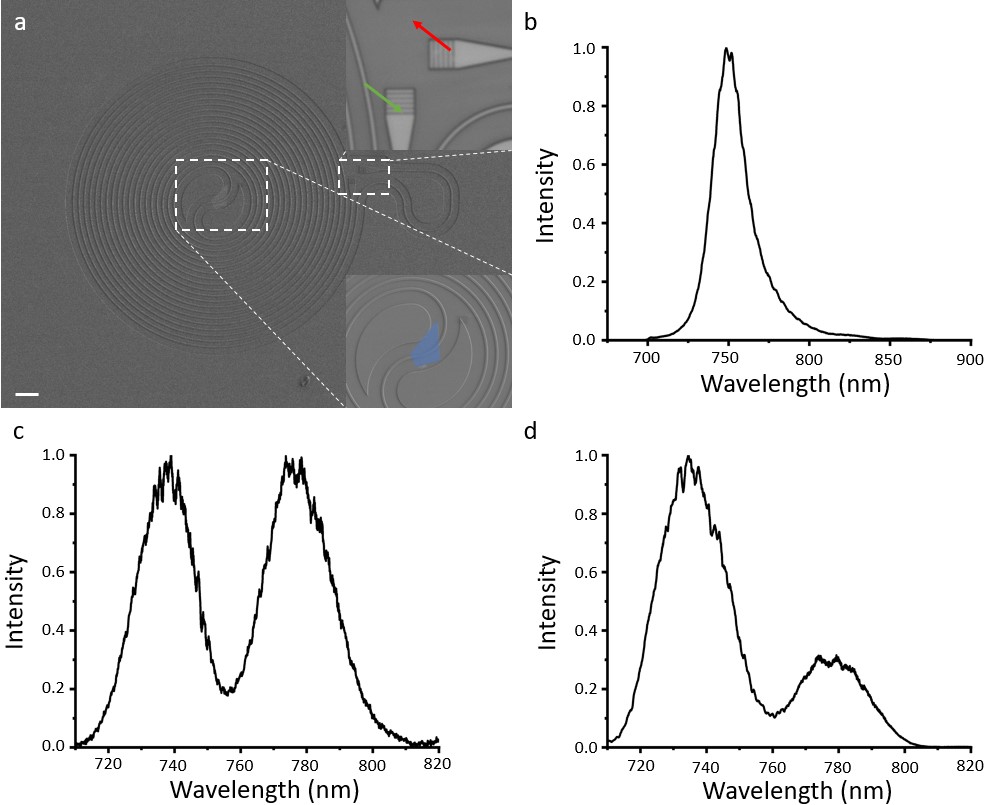}
	\caption{a) SEM image of a silicon nitride spiral. The bottom right inset is a false-color SEM of the integrated monolayer $\text{WSe}_2$. The top right inset is the grating couplers (green - excitation, red - collection). Scale bar is 10 \si{\micro \meter}. b) Room-temperature PL of the monolayer $\text{WSe}_2$ integrated onto the silicon nitride spiral. c) Transmission spectrum for the silicon nitride spiral. d) Transmission spectrum for the silicon nitride spiral with the integrated monolayer $\text{WSe}_2$.}
	\label{fig:spiral2}
\end{figure}

\section{Experimental results and discussion}

We first demonstrate the integration of $\text{WSe}_2$ onto a non-resonant nanophotonic device - a large-area silicon nitride (SiN) spiral (Fig. \ref{fig:spiral2}(a)). Due to its large area, the transmission spectrum is known to be sensitive to contaminants \cite{lee_ultra-low-loss_2012}. Then, resonant nanophotonic devices are demonstrated by the dual integration of two different semiconductor monolayers ($\text{WSe}_2$, $\text{MoSe}_2$) onto neighboring SiN ring resonators. As the two monolayers are integrated in separate transfer steps the samples can be integrated as a heterostructure or onto separate devices depending on the desired experiment. 

We fabricated the underlying nanophotonic devices using a 220 \si{\nano \meter} thick SiN membrane grown via LPCVD on 4 \si{\micro \meter} of thermal oxide on silicon. The samples were obtained from commercial vendor Rogue Valley Microelectronics. We spun roughly 400 \si{\nano \meter} of Zeon ZEP520A, which was coated with a thin layer of Pt/Au that served as a charging layer. The resist was then patterned using a JEOL JBX6300FX electron-beam lithography system with an accelerating voltage of 100 \si{\kilo \volt}. The pattern was transferred to the SiN using a reactive ion etch (RIE) in $\text{CHF}_{3} \text{/O}_{2}$ chemistry. 

Photoluminescence (PL) measurements \cite{tonndorf_photoluminescence_2013} are conducted by exciting the monolayers with a 632 nm HeNe laser. The resulting emission is collected with a free-space confocal microscopy setup and measured in a spectrometer. The spectrometer is a Princeton Instruments IsoPlane SCT-320 Imaging Spectrograph. The transmission spectrum is measured by exciting a grating coupler with a supercontinuum laser (Fianium WhiteLase Micro) and collecting from the other grating coupler (Fig. \ref{fig:spiral2}(a), top right inset). For cavity-coupled PL \cite{ye_monolayer_2015} the sample is directly excited with the HeNe laser and the resulting emission is collected from a grating coupler using a pinhole in the image plane of the confocal microscope. To obtain high signal-to-noise ratio PL we cool down the sample to 80K using liquid nitrogen in a continuous flow cryostat (Janis ST-500).

\begin{figure}
	\centering
	\includegraphics[width=1.0\linewidth]{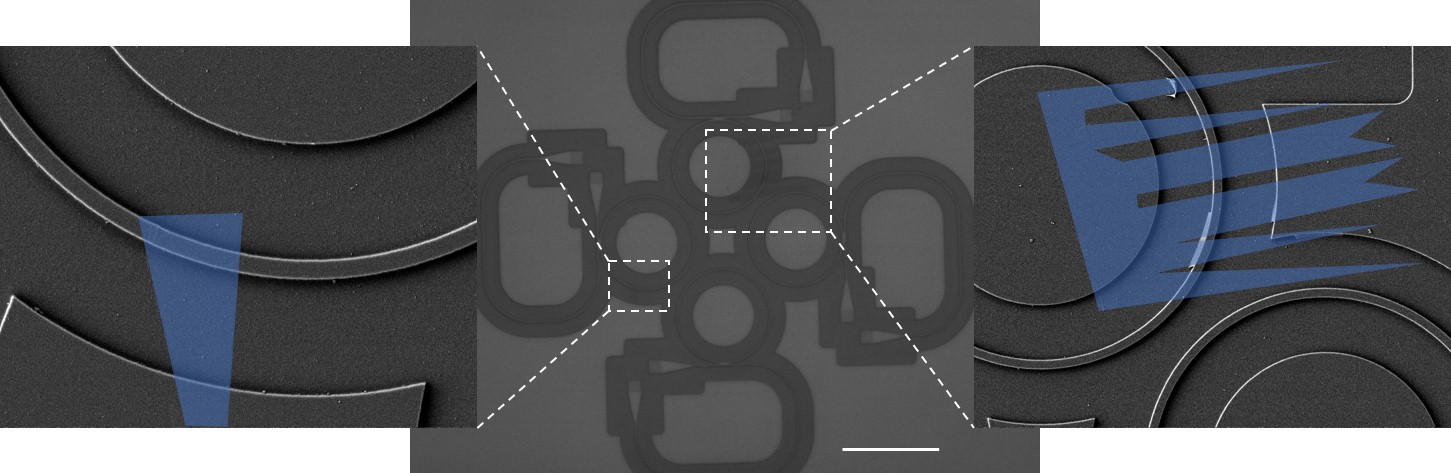}
	\caption{Optical image of exfoliated $\text{WSe}_2$ and $\text{MoSe}_2$ monolayers integrated onto the left and top ring resonators, respectively, with false-color SEM images of the integrated monolayers. Scale bar is 10 \si{\micro \meter}.}
	\label{fig:dualoptical}
\end{figure}

The room-temperature PL with a strong excitonic peak of the $\text{WSe}_2$ monolayer integrated onto the SiN spiral (Fig. \ref{fig:spiral2}(b)) establishes the presence of the vdW material on the waveguide \cite{chernikov_exciton_2014}. The primary peak is attributed to neutral exciton emission, which is indicative of a direct bandgap, semiconducting material when the TMDC vdW materials are exfoliated as monolayers. The secondary sidebands could be due to defects or trions \cite{chow_defect-induced_2015,sidler_fermi_2017}. The before and after transmission spectrum (Fig. \ref{fig:spiral2}(c) and Fig. \ref{fig:spiral2}(d), respectively) for the SiN spiral waveguide integrated with the monolayer $\text{WSe}_2$ illustrates the contamination-free nature of the transfer process. Significant contamination would prevent any transmission spectrum from being measured. The envelope modulation of the spectrum is due to the frequency-dependent coupling efficiency of the grating couplers. The relative amplitude change between the two features in the spectrum is likely due to the angular dependence of the grating couplers. As the measurement is done before and after the transfer - which requires removing the sample from the optical setup - the angular alignment of the confocal microscope objective to the grating coupler will be slightly different \cite{chrostowski_silicon_2015}.

\begin{figure}
	\centering
	\includegraphics[width=1.0\linewidth]{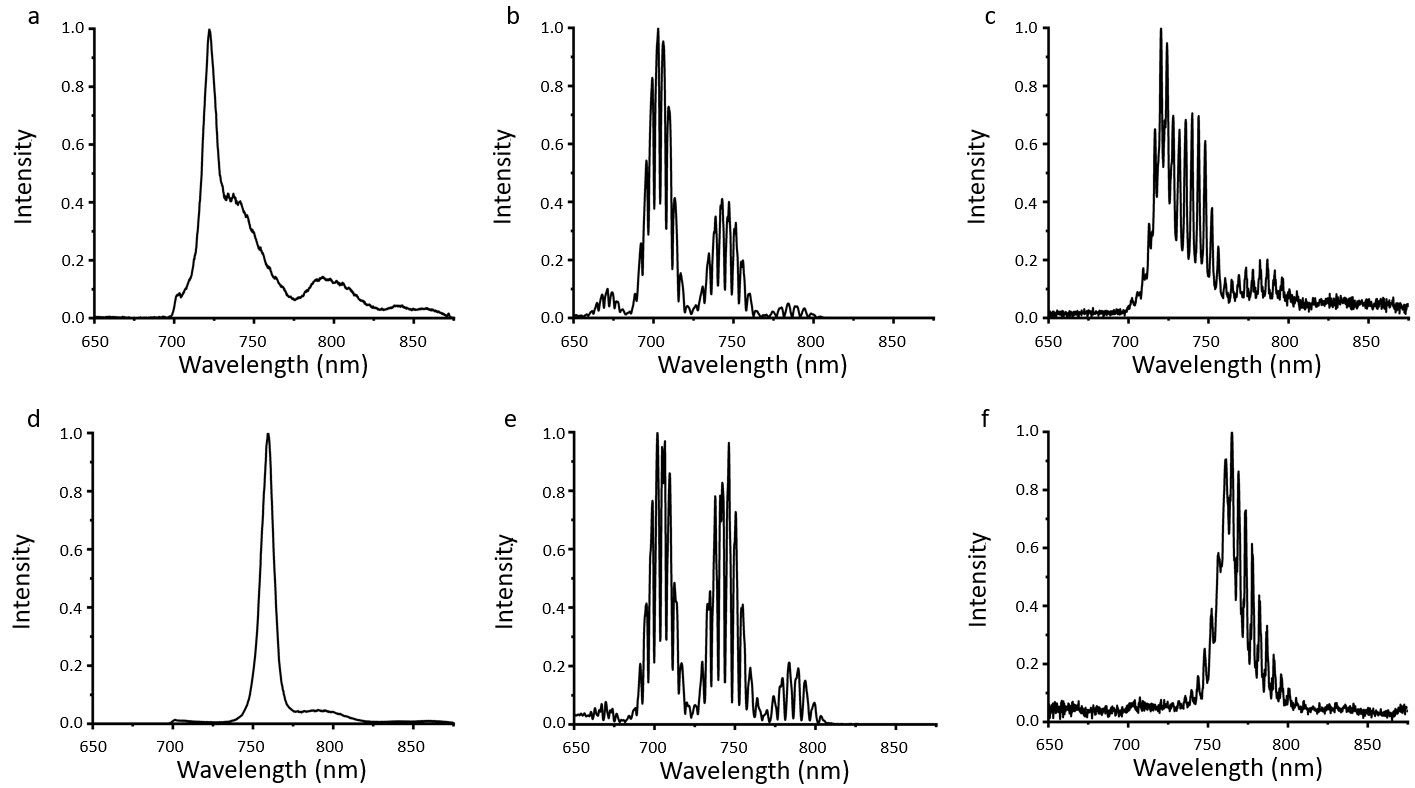}	
	\caption{a) PL of the integrated $\text{WSe}_2$ monolayer at 80 K. b) Transmission spectrum for the ring resonator with the integrated monolayer $\text{WSe}_2$. c) Cavity-coupled PL of the integrated monolayer $\text{WSe}_2$ on the ring resonator. d) PL of the integrated $\text{MoSe}_2$ monolayer at 80 K. e) Transmission spectrum for the ring resonator with the integrated monolayer $\text{MoSe}_2$. f) Cavity-coupled PL of the integrated monolayer $\text{MoSe}_2$ on the ring resonator. The baseline below the cavity resonance peaks in c and f is due to background PL.}
	\label{fig:dualtransfer}
\end{figure}

The method can be extended to integrate vdW materials to disjoint but proximate vdW material nanophotonic devices (Fig. \ref{fig:dualoptical}). The four SiN ring resonators are each separated by 1 \si{\micro \meter} to ensure no coupling between cavities. Each cavity can be independently addressed by input and output grating couplers. Again, the PL of the $\text{WSe}_2$ and $\text{MoSe}_2$ (Fig. \ref{fig:dualtransfer}(a) and Fig. \ref{fig:dualtransfer}(d), respectively) establishes the presence of the monolayers. The low-temperature transmission spectrum for the ring resonators (Fig. \ref{fig:dualtransfer}(b) and Fig. \ref{fig:dualtransfer}(e)) with the integrated monolayers illustrates a contamination-free transfer. The dips in transmission correspond to the resonance in the ring resonators. The separation between the modes corresponds to the free spectral range of the ring resonator. The PL of the $\text{WSe}_2$ and $\text{MoSe}_2$ coupled to the evanescent field of the ring resonators collected from the grating coupler (Fig. \ref{fig:dualtransfer}(c) and Fig. \ref{fig:dualtransfer}(f), respectively) is amplified at the cavity resonances.

\section{Conclusion}

We have presented a method to facilitate the integration of vdW materials onto nanophotonic devices that require minimal contamination from bulk material. A PL measurement is used to identity the presence of vdW materials on the nanophotonic devices. The transmission spectrum of the SiN spiral integrated with a monolayer material demonstrates the contamination-free nature of the described transfer method. The integration of two different transition metal dichalcogenide monolayers onto neighboring SiN ring resonators demonstrates the capability to manually scale the fabrication of devices for rapid prototyping. Our local transfer technique can potentially enable a lithographically defined quantum emitter \cite{kroemer_speculations_2003, wei_size-tunable_2017} deterministically integrated onto a nanocavity, which can reach the few-photon nonlinear optical regime \cite{ryou_strong_2018, delteil_towards_2019, munoz-matutano_emergence_2019} for applications in neuromorphic photonics \cite{shen_deep_2017, lima_machine_2019} and quantum many-body simulation \cite{jiang_photonic_2014, wang_quantum_2017}.

A related local transfer method has been demonstrated by Hemnani et al \cite{hemnani_2d_2018}. The primary advantages of our method compared to that technique are the backward compatibility with the standard dome transfer method (5-10 additional minutes) and the opportunity for polymer-free heterostructures. Additionally, our placement precision is within the width of a waveguide, or $\pm$ 0.5 \si{\micro \meter}. This spatial resolution is almost an order of magnitude larger than the one reported in Hemnani et al \cite{hemnani_2d_2018}. Finally, our work conclusively proved local transfer of a monolayer transition metal dichalcogenide.

\section*{Acknowledgments}
The research was supported by NSF-1845009, NSF-ECCS-1708579 and AFOSR grant FA9550-17-C-0017 (Program Manager Dr. Gernot Pomrenke). Part of this work was conducted at the Washington Nanofabrication Facility / Molecular Analysis Facility, a National Nanotechnology Coordinated Infrastructure (NNCI) site at the University of Washington, which is supported in part by funds from the National Science Foundation (awards NNCI-1542101, 1337840 and 0335765), the National Institutes of Health, the Molecular Engineering \& Sciences Institute, the Clean Energy Institute, the Washington Research Foundation, the M. J. Murdock Charitable Trust, Altatech, ClassOne Technology, GCE Market, Google and SPTS. A.R. acknowledges support from the IC Postdoctoral Research Fellowship.

\section*{Disclosures}
The authors declare no conflicts of interest.

\bibliography{references}
\bibliographystyle{unsrtnat}

\end{document}